\documentclass[12pt]{article}
\usepackage{amssymb,amsmath}

\catcode`\@=11

\renewcommand{\subsection}{\@startsection%
{subsection}{2}{0mm}{-\baselineskip}{0.5\baselineskip}%
{\normalfont\normalsize\itshape}}

\@addtoreset{equation}{section}
\def\theequation{\thesection\arabic{equation}}

\def\@normalsize{\@setsize\normalsize{15pt}\xiipt\@xiipt
\abovedisplayskip 14pt plus3pt minus3pt%
\belowdisplayskip \abovedisplayskip
\abovedisplayshortskip  \z@ plus3pt%
\belowdisplayshortskip  7pt plus3.5pt minus0pt}
\def\small{\@setsize\small{13.6pt}\xipt\@xipt
\abovedisplayskip 13pt plus3pt minus3pt%
\belowdisplayskip \abovedisplayskip
\abovedisplayshortskip  \z@ plus3pt%
\belowdisplayshortskip  7pt plus3.5pt minus0pt
\def\@listi{\parsep 4.5pt plus 2pt minus 1pt
            \itemsep \parsep
            \topsep 9pt plus 3pt minus 3pt}}

\def\underline#1{\relax\ifmmode\@@underline#1\else
        $\@@underline{\hbox{#1}}$\relax\fi}
\@twosidetrue
\relax

\catcode`@=12

\catcode`\@=11

\def\section{\@startsection{section}{1}{\z@}{3.5ex plus 1ex minus
   .2ex}{2.3ex plus .2ex}{\large\bf}}
\def\thesection{\arabic{section}.}

\def\ps@headings{\def\@oddfoot{}\def\@evenfoot{}
\def\@oddhead{\hbox{}\hfill
        \makebox[.5\textwidth]{\raggedright\ignorespaces --\thepage{}--
        \hfill }}
\def\@evenhead{\@oddhead}
\def\subsectionmark##1{\markboth{##1}{}}
}

\ps@headings

\catcode`\@=12

\relax
\def\r#1{\ignorespaces $^{#1}$}
\def\figcap{\section*{Figure Captions\markboth
        {FIGURECAPTIONS}{FIGURECAPTIONS}}\list
        {Fig. \arabic{enumi}:\hfill}{\settowidth\labelwidth{Fig. 999:}
        \leftmargin\labelwidth
        \advance\leftmargin\labelsep\usecounter{enumi}}}
 \relax
\def\tablecap{\section*{Table Captions\markboth
        {TABLECAPTIONS}{TABLECAPTIONS}}\list
        {Table \arabic{enumi}:\hfill}{\settowidth\labelwidth{Table 999:}
        \leftmargin\labelwidth
        \advance\leftmargin\labelsep\usecounter{enumi}}}
 \relax
\def\reflist{\section*{References\markboth
        {REFLIST}{REFLIST}}\list
        {[\arabic{enumi}]\hfill}{\settowidth\labelwidth{[999]}
        \leftmargin\labelwidth
        \advance\leftmargin\labelsep\usecounter{enumi}}}
 \relax

\catcode`\@=11

\def\marginnote#1{}
\newcount\hour
\newcount\minute
\newtoks\amorpm
\hour=\time\divide\hour by60
\minute=\time{\multiply\hour by60 \global\advance\minute by-
\hour}
\edef\standardtime{{\ifnum\hour<12 \global\amorpm={am}%
    \else\global\amorpm={pm}\advance\hour by-12 \fi
    \ifnum\hour=0 \hour=12 \fi
    \number\hour:\ifnum\minute<100\fi\number\minute\the\amorpm}}
\edef\militarytime{\number\hour:\ifnum\minute<100\fi\number\minute}
\def\draftlabel#1{{\@bsphack\if@filesw {\let\thepage\relax
  \xdef\@gtempa{\write\@auxout{\string
    \newlabel{#1}{{\@currentlabel}{\thepage}}}}}\@gtempa
    \if@nobreak \ifvmode\nobreak\fi\fi\fi\@esphack}
     \gdef\@eqnlabel{#1}}
\def\@eqnlabel{}
\def\@vacuum{}
\def\draftmarginnote#1{\marginpar{\raggedright\scriptsize\tt#1}}
\def\draft{\oddsidemargin -.5truein
        \def\@oddfoot{\sl preliminary draft \hfil
        \rm\thepage\hfil\sl\today\quad\militarytime}
        \let\@evenfoot\@oddfoot \overfullrule 3pt
        \let\label=\draftlabel
        \let\marginnote=\draftmarginnote
   
\def\@eqnnum{(\theequation)\rlap{\kern\marginparsep\tt\@eqnlabel}%
\global\let\@eqnlabel\@vacuum}  }
\def\preprint{\twocolumn\sloppy\flushbottom\parindent 1em
        \leftmargini 2em\leftmarginv .5em\leftmarginvi .5em
        \oddsidemargin -.5in    \evensidemargin -.5in
        \columnsep 15mm \footheight 0pt
        \textwidth 250mmin      \topmargin  -.4in
        \headheight 12pt \topskip .4in
        \textheight 175mm
        \footskip 0pt
        
\def\@oddhead{\thepage\hfil\addtocounter{page}{1}\thepage}
        \let\@evenhead\@oddhead \def\@oddfoot{} \def\@evenfoot{}
}
\def\titlepage{\@restonecolfalse\if@twocolumn\@restonecoltrue\onecolumn
     \else \newpage \fi \thispagestyle{empty}\c@page\z@
        \def\thefootnote{\fnsymbol{footnote}} }
\def\endtitlepage{\if@restonecol\twocolumn \else  \fi
        \def\thefootnote{\arabic{footnote}}
        \setcounter{footnote}{0}}  
\catcode`@=12
\relax

\def\ps@headings{\def\@oddfoot{}\def\@evenfoot{}
\def\@oddhead{\hbox{}\hfill
        \makebox[.5\textwidth]{\raggedright\ignorespaces --\thepage{}--
        \hfill }}
\def\@evenhead{\@oddhead}
\def\subsectionmark##1{\markboth{##1}{}}
}

\ps@headings

\relax

\def\firstpage#1#2#3#4#5#6{
\begin{document}
\begin{titlepage}
\nopagebreak
\title{\begin{flushright}
        \vspace*{-1.8in}
        {\normalsize CERN--TH/97-357}\\[-8mm]
        {\normalsize CPTH--S586.1297}\\[-8mm]
        {\normalsize RI--12--97}\\[-8mm]
        {\normalsize hep-th/9712084} \\[4mm]
\end{flushright}
\vfill
\vskip -.5cm
{#3}}
\author{\large #4 \\[.5cm] #5}
\maketitle
\vskip -7mm
\nopagebreak
\begin{abstract}
{\noindent #6}
\end{abstract}
\vfill
\begin{flushleft}
\rule{16.1cm}{0.2mm}\\[-3mm]
$^{\dagger}${\small Laboratoire Propre du CNRS UPR A.0014.}\\
CERN--TH/97-357,\ CPTH--S586.1297, RI--12--97\\
December 1997, revised April 1998
\end{flushleft}
\thispagestyle{empty}
\end{titlepage}}

%
%
\newcommand{\ie}{\hbox{\it i.e.}\ }
\newcommand{\Rank}{{\rm Rank}\ }
\newcommand{\Tr}{{\rm Tr}\ }
\newcommand{\STr}{{\rm Str}\ }
\newcommand{\axion}{\hbox{\Large $a$} }
\newcommand{\Zint}{{\mbox{\sf Z\hspace{-3.2mm} Z}}}
\newcommand{\A}{{\cal A}}
\newcommand{\C}{{\cal C}}
\newcommand{\E}{{\cal E}}
\newcommand{\G}{{\cal G}}
\newcommand{\K}{{\cal K}}
\newcommand{\M}{{\cal M}}
\newcommand{\R}{{\cal R}}
\newcommand{\tg}{\tilde g}
\def\a{\alpha}
\def\m{\mu}
\def\n{\nu}
\def\r{\rho}
\newcommand{\Real}{{\mbox{I\hspace{-2.2mm} R}}}
\renewcommand{\sp}{\; , \; \; }

\date{}
\firstpage{3155}{}
{\large\sc M-Theory and U-duality on $T^d$ with Gauge Backgrounds}
{N.A. Obers$^a$, B. Pioline$^{\,a,b}$ and E. Rabinovici$^{a,c}$}  
{\normalsize\sl
\normalsize\sl $^a$Theory Division, CERN, 1211 Geneva 23,
Switzerland\\[-3mm]
\normalsize\sl$^b$Centre de Physique Th{\'e}orique,
Ecole Polytechnique,$^\dagger$
{}F-91128 Palaiseau, France\\[-3mm]
\normalsize\sl$^c$Racah Institute of Physics, The Hebrew University, Jerusalem
91904, Israel
}
{The full U-duality symmetry of toroidally compactified
M-theory can only be displayed by allowing non-rectangular tori
with expectation values of the gauge fields.
We construct an $E_d(\Zint)$ U-duality invariant mass formula
incorporating non-vanishing gauge backgrounds
of the M-theory three-form $\C$. We interpret this mass
formula from the point of view of the Matrix gauge theory,
and identify the coupling of the three-form to the gauge theory
as a topological theta term, in agreement with earlier
conjectures. We give a derivation of this fact from D-brane
analysis, and obtain the Matrix gauge theory description
of other gauge  backgrounds allowed by the Discrete
Light-Cone Quantization.
We further show that the conjectured extended U-duality symmetry
of  Matrix theory on $T^d$ in the Discrete Light-Cone Quantization
has an implementation as an action of
 $E_{d+1}(\Zint)$ on the BPS spectrum.
Some implications for the proper interpretation of the rank $N$ of the Matrix
gauge theory are discussed.
}

\eject
\section{Introduction}
It was suggested that toroidal compactification of M-theory \cite{wit}
in the infinite momentum frame
is described by a Matrix gauge theory on the T-dual torus
\cite{bfss,taylor,sdual}. This gauge theory ought to reduce
to Supersymmetric Yang-Mills (SYM) theory
with 16 supercharges for up to three compact directions.
When more directions are compactified, several suggestions
have been made on how to supplement SYM
with new degrees of freedom at short distances, still avoiding
the coupling to gravity \cite{rozali,6dims}. Non-perturbative
dualities of this supersymmetric gauge theory, together with
the mapping class group of the torus on which the gauge theory lives,
account for the U-dualities \cite{ht}  of the corresponding maximally
supersymmetric type II theories. For instance, the U-duality
groups $Sl(2,\Zint)\times Sl(3,\Zint)$ of M-theory compactified
on a three-torus (type II compactified on a two-torus)
correspond to the electric-magnetic duality of SYM in 1+3
dimensions and the reparametrizations of the dual three-torus
respectively \cite{sdual} (see \cite{aharony} for a relation of
this duality to the membrane-fivebrane duality).

For higher dimensional compactifications, it has been shown
that the electric-magnetic duality $\Zint_2$ on
the $T^3\subset T^d$ fibres,
together with the permutations ${\cal S}_d$
of the torus directions, generates a finite group, the Weyl group\footnote{
We denote by $G\ltimes G'$ the group obtained by taking
the generators of $G$ and $G'$ together.}
${\cal W}(E_d)=\Zint_2\ltimes {\cal S}_d$ of the Cremmer-Julia
hidden symmetry $E_d(\Real)$ \cite{egkr}. This Weyl group is the subgroup
of U-duality preserving the rectangular shape of the torus
and the vanishing expectation value of the M-theory gauge
three-form $\C_{IJK}$.
The U-duality group includes $Sl(d,\Zint)\ltimes SO(d-1,d-1,\Zint)$,
corresponding to the mapping class group of the M-theory compactification
together with the perturbative string T-duality;
in the following, we shall refer to this product as $E_d(\Zint)$,
although it is not known whether this subgroup is sufficient\footnote{We thank
B. Julia for pointing this out.}
 to generate
the U-duality group $E_d(\Real)\bigcap Sp(28,\Zint)$ conjectured by
Hull and Townsend \cite{ht}  when $d\le 7$.
In particular, $E_d(\Zint)$ contains
elements shifting the three-form potential $\C_{IJK}$ by
an integer, or acting as modular transformations on the torus.
The full U-duality symmetry can only be displayed by allowing such
skew tori with arbitrary uniform value of the gauge potential.

In Section 2, we will extend the analysis of Ref.\ \cite{egkr}
to determine
 U-duality invariant mass formulae for the 1/2 BPS states of
M-theory
 compactified on general tori with non-vanishing gauge background.
Our strategy will be to first construct a T-duality invariant mass formula
valid in the presence of an arbitrary $B_{ij } =\C_{sij}$ field,  derive
the action of the spectral flow $B_{ij} \rightarrow B_{ij} + \Delta B_{ij}$,
and subsequently covariantize this flow
to include the generators $\C_{IJK} \rightarrow \C_{IJK}
+ \Delta \C_{IJK}$ as well as the additional Borel generators that appear for
$d \geq 6$.
We will point out the relation with results previously obtained in the
context of instanton corrections to type II string theory couplings
\cite{pk}.

The Matrix gauge theory corresponding to compactification
on skew tori is simply a supersymmetric gauge theory defined
on the dual skew tori. As already proposed in Ref.\ \cite{bcd},
the expectation value of the gauge potential on the other hand
turns into a set of topological couplings on the dual torus.
In the particular example of M-theory on a three-torus, this
is simply the theta-angle $\theta=\C_{123}$, which extends the
electric-magnetic duality from $\Zint_2$ to $Sl(2,\Zint)$.
These couplings have, however, been inferred rather than derived, and a
recent argument \cite{sen,seiberg} allows a more systematic
derivation from the D-brane action, at least for M-theory backgrounds
corresponding to Ramond-Ramond (RR) potentials in the type IIA
string description. In Section 3, we will translate the M-theory
mass formula into the Matrix gauge theory language and show the
agreement with the coupling  derived from the D-brane analysis.

M(atrix) theory still lacks a proof of eleven-dimensional Lorentz
covariance to shorten its name to M-theory. In the original
conjecture \cite{bfss}, this feature was credited to the large-$N$
infinite-momentum limit. The much stronger Discrete Light
Cone (DLC) conjecture \cite{dlcq}, if correct, allows
Lorentz invariance to be checked at finite $N$ -- or rather
at finite $N$'s, since the non-manifest Lorentz generators mix
distinct $N$ superselection sectors. In particular, M(atrix) theory
on $T^d$ in the DLC should exhibit a U-duality $E_{d+1}(\Zint)$,
if one assumes that U-duality is unaffected by light-like
compactifications. In Section 4, we shall show that
the promotion of the rank $N$ to an ordinary charge \cite{hv} allows
the existence of an $E_{d+1}(\Zint)$ action on the spectrum of BPS states.
Related results have been obtained in Refs.
\cite{cds,hull,blau,hull2,dewit}.

\section{M-theory BPS states and Invariant Mass Formulae}

The authors of Ref.\ \cite{egkr} have investigated the ${\cal W}(E_d)$ orbits
of two BPS states that are required to exist in the Matrix
gauge theory reducing to SYM in the infrared\footnote{We only
deviate from the notations used in Ref.\ \cite{egkr}
in that the light-cone compact radius $R_{11}$ is now $R_l$; $l_p$
is therefore the eleven-dimensional Planck length, $R_I$ the
radii of the compactification torus, $s_I$ the radii of the
gauge theory torus, $V_s$ its volume, $g$ the gauge coupling.}
: the
quantum of {\it flux}, with energy $P^-_F=g^2 s_I^2 / (N V_s)$,
and the quantum of {\it momentum}, with energy $P^-_{M}=1/s_I$.
{}From the M-theory point of view, they correspond to a
Kaluza-Klein excitation with mass $\M_F=\sqrt{P^+ P^-}=1/R_I$,
and to a membrane wrapped on a circle of the torus times
the light-cone direction, yielding a particle with mass
$\M_M=R_l R_I / l_p^3$. The generalization to skew tori
is immediate:
\begin{equation}
\M_F^2 = m_I g^{IJ} m_J
\sp
\M_M^2 = \frac{R_l^2}{l_p^6} n^I g_{IJ} n^J \;.
\end{equation}
Here $m_I$ describes the KK momentum, while $n^I$ labels
the cycle of $T^d$ on which the membrane wraps.
In the following, we shall describe how these mass formulae
can be extended to include all states of these two U-duality multiplets
and the dependence on all M-theory moduli.

\subsection{The flux multiplet}
Under electric-magnetic duality on three of the
directions of the Yang-Mills torus, it has been shown that
the flux quantum turns into a set of states with masses
$$
\frac{1}{R_I}\sp
\frac{R_I R_J}{l_p^3}\sp
\frac{R_I R_J R_K R_L R_M}{l_p^6}\sp
\frac{R_I^2 R_J R_K R_L R_M R_N R_P}{l_p^9}\sp
\frac{R_I^2 R_J^2 R_K^2 R_L R_M R_N R_P R_Q}{l_p^{12}}\sp
$$
$$
\frac{R_I^2 R_J^2 R_K^2 R_L^2 R_M^2 R_N^2 R_P R_Q}{l_p^{15}}\sp
\frac{R_I^3 R_J^2 R_K^2 R_L^2 R_M^2 R_N^2 R_P^2 R_Q^2}{l_p^{18}}
\; , $$
starting to appear for $d=1,2,5,7,8,8,8$ respectively (indices
$I,J$, etc., are distinct). The charges labelling superposition
of these states can therefore be cast into integer tensors
$m_I,m^{IJ},m^{IJKLM},m^{I;JKLMNPQ},m^{IJK;LMNPQRST}$, etc. where
the groups of indices separated by a semi-colon are antisymmetric
and no symmetry accross
a semi-colon is assumed. In short, the flux multiplet is
described by a set of integer charges
$$m_1\ ,\ m^2,\ m^5,\ m^{1;7},\ m^{3;8},\ m^{6;8},\ m^{1;8;8},$$
where the integers label the number of indices.
This yields the correct number of
charges to make up the representations of $E_d$ U-duality groups.
The contribution of a given
charge tensor to the total square mass is simply given by its
square norm induced by the torus metric $g_{IJ}$,
with the appropriate symmetry factor and power of $l_p$:
\begin{equation}
\label{mfsl}
\M_F^2 = m_I g^{IJ} m_J + \frac{1}{2!~l_p^6} m^{IJ} g_{IK} g_{JL} m^{KL}
+\frac{1}{5!~l_p^{12}} m^{IJKLM} g_{IN} g_{JP} g_{KQ} g_{LR} g_{MS}
m^{NPQRS} +\dots
\end{equation}

The mass formula is compatible with the interpretation of
the $m_I$ charge as the KK momentum along the $I$-th
direction of the transverse torus, $m^{IJ}$ as the wrapping number of the
M-theory membrane on
a two-cycle of the same torus,
and $m^{IJKLM}$ as the wrapping number of the M-theory five-brane on
a five-cycle. The charge $m^{1;7}$ yields a tension of the form
$R_I^2/l_p^9$,  corresponding to Taub-NUT gravitational
monopole on the $R_I$ direction.
The higher charges are not understood at present.
As in Ref.\ \cite{egkr} we draw consequences from symmetry arguments
in the hope that dynamical issues will be resolved.

\subsection*{T-duality invariant mass formulae}

As it stands, the mass formula (\ref{mfsl}) is invariant under $Sl(d,\Zint)$,
but not under $SO(d-1,d-1,\Zint)$ T-duality: it only holds when the
background gauge fields vanish. In order
to reinstate the dependence on $\C_{IJK}$,
we first decompose the flux multiplet
as a sum of T-duality irreducible representations,
and couple them to the NS two-form $B_{ij}=\C_{sij}$.
For that purpose, we choose a direction $s\in\{1,\dots,d\}$ on $T^d$
and rewrite the mass formula (\ref{mfsl}) in terms of the type IIA
string theory variables $(g_s,l_s)$, related to the
M-theory variables by $R_{s}=l_s g_s, l_p=l_s g_s^{1/3}$:
\begin{equation}
\label{gdec}
\begin{split}
\M_F^2 =& \left[ \frac{m_s^2}{g_s^2}+ (m_1)^2 \right]
+\left[ (m^{s1})^2 + \frac{(m^{2}) ^2}{g_s^2} \right]
+\left[ \frac{(m^{s4})^2}{g_s^2} + \frac{(m^{5})^2}{g_s^4} \right]
\\
&+\left[ \frac{(m^{s;s6})^2}{g_s^2} +
        \frac{(m^{s;7})^2+(m^{1;s6})^2}{g_s^4}+
        \frac{(m^{1;7})^2}{g_s^6} \right]
\\
&+\left[ \frac{(m^{s2;s7})^2}{g_s^4} +
        \frac{(m^{3;s7})^2}{g_s^6} \right]
+\left[ \frac{(m^{s5;s7})^2}{g_s^6} +
        \frac{(m^{6;s7})^2}{g_s^8} \right]
+\left[ \frac{(m^{s;s7;s7})^2}{g_s^6} + \frac{(m^{1;s7;s7})^2}{g_s^8} \right]
\;,
\end{split}
\end{equation}
where we retained only the powers of the
string coupling and the index structure.
T-duality commutes with the grading in powers of $g_s$, so
we learn that the flux multiplet decomposes as a sum of
five representations:
$$V=(m_1,m^{s1})\sp S_B=(m_s,m^2,m^{s4},m^{s;s6}) \sp
T=(m^5,m^{1;s6},m^{s;7},m^{s2;s7}),$$
$$S_A=(m^{1;7},m^{3;s7},m^{s5;s7},m^{s;s7;s7})\sp
V'=(m^{6;s7},m^{1;s7;s7}) \; .$$
The irrep $V$ is merely a vectorial representation of $SO(d-1,d-1,\Zint)$,
for which the mass formula is known from the the usual tori partition
functions \cite{gpr}:
\begin{equation}
\label{mv}
\M_V^2= \left( m_i +  B_{ji} m^{sj}  \right) g^{ik}
\left( m_k + B_{lk} m^{sl}  \right) + m^{si} g_{ij} m^{sj} \; .
\end{equation}
The irrep $S_B$ on the other hand already arose in Ref.\ \cite{pk}
as the set of type IIB D-brane charges. It is well known that
the type II RR gauge fields transform as a spinorial representation
of $SO(d-1,d-1,\Real)$, the Clifford algebra being generated by
inner and wedge products with the torus first cohomology,
and the chirality depending on the type A or B \cite{pk}. The corresponding
charges therefore transform as a (conjugate) spinor, hence the
notation $S_B$. Note that this does not imply that the states in
$S_B$ correspond to the D-branes of type II string theory, but simply that they
transform in the same way. The T-duality invariant mass formula comes as
a by-product of the analysis of Ref.\ \cite{pk}:
\begin{equation}
\label{msb}
\begin{split}
\M_{S_B}^2=&
\left(m_s+\frac{1}{2} B_{ij} m^{ij}+\frac{1}{2\cdot 2^2} B_{ij} B_{kl} m^{sijkl}
+ \frac{1}{3! \cdot 2^3} B_{ij} B_{kl} B_{mn} m^{s;sijklmn} +\dots
 \right)^2\\
&+\frac{1}{2}\left(m^{ij}+\frac{1}{2}B_{kl} m^{sklij}+
\frac{1}{2\cdot 2^2}B_{kl} B_{mn} m^{s;sklmnij}+
\dots \right)^2\\
&+\frac{1}{4!}\left(m^{sijkl}+\frac{1}{2}B_{mn} m^{s;smnijkl}+\dots\right)^2
+\frac{1}{6!}\left(m^{s;sijklmn}+\dots\right)^2 +\dots
\end{split}
\end{equation}
In the above equation, we have again dropped the metric contractions
and the powers of $l_s$. The dots include the higher even forms
arising in the reduction of the spinor of $SO(d-1,d-1,\Real)$ to
antisymmetric forms of $Sl(d-1,\Real)$, but are irrelevant for $d\le 8$.
The representation $T$ reduces to a singlet at $d-1=5$,
when it starts appearing, and to
a vector $V$ when $d-1=6$ (upon dualization of the {\bf 5} and {\bf 6}
indices). For $d-1=7$, it extends to an $SO(d-1,d-1,\Real)$ two-form
together with a singlet,
as is easily seen by dualizing on $T^7$ to
$(m_2,m^{1s}_1,m^{s},m^{s2;s})$. The mass formula is then
obtained\footnote{This requires a precise identification of the
T-duality singlet among $m^s$ and ${\rm Tr}\,m_1^{1s}$.}
by tensor product from Eq.\ (\ref{mv}):
\begin{equation}
\label{mas}
\begin{split}
\M_{T}^2=&
\frac{1}{5!}
\left(m^{ijklm}+B_{np} \left( \frac{1}{2} m^{s;npijklm} -m^{n;spijklm}
\right) +
\frac{1}{2} B_{np} B_{qr} m^{snq;sprijklm}
\right)^2\\
&+\frac{1}{6!}\left( m^{p;sijklmn}- B_{qr} m^{sqp;srijklmn}\right)^2
+ \frac{1}{7!} \left( m^{s;ijklmnp} + \frac{1}{2} B_{qr} m^{sqr;sijklmnp}
 \right)^2 \\
&+\frac{1}{2\cdot 7!}\left(m^{sqr;sijklmnp}\right)^2 \; .
\end{split}
\end{equation}
Finally, the irreps $S_A$ and $V'$ only arise for $d-1=7$.
$S_A$ is, after dualizing
the {\bf 7} indices, a sum of odd forms of $Sl(d-1,\Real)$, and therefore
a spinor representation of $SO(d-1,d-1,\Zint)$ with chirality opposite
to $S_B$:
\begin{equation}
\label{msa}
\begin{split}
\M_{S_A}^2=&
\left(m^{i;7}+\frac{1}{2}B_{jk} m^{jki;s7}
+\frac{1}{8} B_{jk} B_{lm} m^{sjklmi;s7}
+ \dots \right)^2\\
&+\frac{1}{3!}\left(m^{ijk;s7}+\frac{1}{2}B_{lm}m^{slmijk;s7}+ \dots \right)^2
+\frac{1}{5!}\left(m^{sijklm;s7}+\dots\right)^2 +
\left(m^{s;s7;s7} \right)^2  +\dots\ ,
\end{split}
\end{equation}
while $V'$ reduces to a representation $V$ after dualizing the
{\bf 6} and {\bf 7} indices.

\subsection*{T-duality spectral flows}
\enlargethispage{2cm}
Adding $\M^2_{ \{ V,S_B,T,S_A,V' \} }$ together, we obtain the T-duality
invariant flux multiplet mass formula, which is still of the form in
Eq.(\ref{gdec})
but for replacing the $m$ charges with shifted charges $\tilde{m}$
incorporating the effect of the $B$ field, e.g.
\begin{equation}
\tilde{m}_s = m_s + \frac{1}{2} B_2 m^2  + \frac{1}{8} B_2^2 m^{s4}
+ \frac{1}{48} B_2^3 m^{s;s6} \ .
\end{equation}
The mass spectrum is thus globally invariant under the
integer shift $B_{ij} \rightarrow B_{ij} + \Delta B_{ij}$, even though
the latter induces a {\it spectral flow} within each
T-duality multiplet:
\begin{equation}
\label{spft}
\begin{array}{ll}
V:   & m_i \rightarrow  m_i + \Delta B_{ji} m^{sj}\ ,\quad
       m^{si} \rightarrow m^{si} \vspace{2mm} \ ,\\
S_B: & m_s \rightarrow m_s + \frac{1}{2}\Delta B_{ij}m^{ij} \ ,\quad
       m^{ij}\rightarrow m^{ij} + \frac{1}{2} \Delta B_{kl} m^{sklij} \ ,\\
     & m^{sijkl}\rightarrow m^{sijkl} + \
                \frac{1}{2} \Delta B_{mn}m^{s;smnijkl} \ ,\quad
       m^{s;sijklmn}\rightarrow m^{s;sijklmn} \ ,\vspace{2mm}\\
T:   & m^{ijklm} \rightarrow m^{ijklm} +   \Delta B_{np} (
  \frac{1}{2} m^{s;npijklm} -       m^{n;spijklm}   )\ ,  \\
     & m^{p;sijklmn} \rightarrow m^{p;sijklmn}-\Delta B_{qr} m^{sqp;srijklmn}
\ ,\\
     & m^{s;ijklmnp} \rightarrow m^{s;ijklmnp} + \frac{1}{2} \Delta B_{qr}
       m^{sqr;sijklmnp}  \ , \\
     &  m^{sqr;sijklmnp} \rightarrow m^{sqr;sijklmnp} \ , \vspace{2mm}\\
S_A: & m^{i;jklmnpq} \rightarrow m^{i;jklmnpq} + \frac{1}{2}\Delta B_{rt}
m^{rti;sjklmnpq} \ ,\quad \\
     & m^{ijk;slmnpqrt}\rightarrow m^{ijk;slmnpqrt} + \frac{1}{2} \Delta B_{uv}m^{suvijk;slmnpqrt}\ , \\
     & m^{sijklm;snpqrtuv}\rightarrow m^{sijklm;snpqrtuv} + \frac{1}{2} \Delta B_{wx} m^{s;swxijklm;snpqrtuv}  \ ,\quad \\
     & m^{s;s7;s7}\rightarrow m^{s;s7;s7} \ ,\vspace{2mm}\\
V':  & m^{ijklmn;spqrtuvw} \rightarrow m^{ijklmn;spqrtuvw} - \frac{1}{2} \Delta B_{xy} m^{x;syijklmn;spqrtuvw} \ ,\quad \\
     & m^{1;s7;s7}\rightarrow m^{1;s7;s7}  \\
\end{array}
\end{equation}
The flow indeed acts as an automorphism on the charge lattice; note that,
except for the highest
weight ($m^{1;8;8}$ in $d=8$), the charges cannot be restricted
to positive integers. This fact will be of use in Section 4.

Alternatively, the above spectral flow can be recast into a system of
differential equations for the shifted charges $\tilde m$, e.g.
\begin{equation}
S_B:
\begin{array}{llllll}
\frac{ \partial \tilde{m}_s }{\partial B_{ij} }  &=&
\frac{1}{2} \tilde m^{ij}\ ,\quad &
 \frac{ \partial \tilde{m}^{ij} }{\partial B_{kl} }  &=&
\frac{1}{2} \tilde m^{sijkl} \ ,\quad\\
\frac{ \partial \tilde{m}^{sijkl} }{\partial B_{mn} }  &=&
\frac{1}{2} \tilde m^{s;sijklmn} \ ,\quad
&
 \frac{ \partial \tilde{m}^{s;sijklmn} }{\partial B_{pq} }  &=& 0  \ ,\quad
\end{array}
\end{equation}
which can be integrated to yield the mass formula; the constants
of integration correspond to the integer charges $m$.
The integrability of this  system of differential equations follows from
the commutativity of the spectral flow.

\subsection{U-duality spectral flows}

The mass formula obtained so far is invariant under T-duality and
holds for vanishing values of RR gauge backgrounds.
In order to obtain a U-duality invariant mass formula, we have
to allow expectation values of the  M-theory gauge three-form $\C_{IJK}$,
which extends the NS two-form $B_{ij} = \C_{sij}$; the expectation value
of the  RR one-form is already incorporated as the off-diagonal component
$\A_i=g_{si}/R_s^2 \ne 0$ of the metric in Eq.(\ref{mfsl}).
 For $d\ge 6$, one should also
allow expectation values of the six-form $\E_{IJKLMN}$ Poincar{\'e}-dual
to $\C_{IJK}$ in eleven dimensions:
in the string theory language,
it corresponds to the RR five-form $\E_{s5}$ together with the
NS six-form dual to $B_{\mu\nu}$ in ten dimensions.
For $d=8$, the eight KK gauge fields $g_{\mu I}$
in three space-time dimensions are dual to eight scalars $\K_I$, which,
together with $g_{IJ},\C_{3}$ and $\E_6$, span the $E_8/SO(16)$ scalar
manifold. $\K_I$ may alternatively be thought of as the form $\K_{1;8}$.
$\K_{s;s7}$ is then nothing but the expectation value of the RR seven-form
on the string theory seven-torus.

Together with the
Teichm{\"u}ller transformations
$\gamma_I \rightarrow \gamma_I + \gamma_J$ on the
cycles $\gamma_I$ of the compactification torus,
the integer shifts of the gauge potential expectation
values provide the necessary {\it Borel} generators
to extend the finite Weyl group ${\cal W}(E_d)$
to the full $E_d(\Zint)$ U-duality group.
These two sets of generators
are actually conjugated under T-duality, since a skew torus
turns into a torus with non-vanishing $B_{ij}$ field under T-duality.

In order to reinstate the $\C_{IJK}$ dependence in the mass formula,
we covariantize the $B_{ij}=\C_{sij}$ spectral flow
(\ref{spft}) under $Sl(d,\Zint)$. This yields
\begin{equation}
\label{spcfl}
\begin{array}{llll}
m_I &\rightarrow& m_I &+ \frac{1}{2} \Delta\C_{JKI}~ m^{JK}\\
m^{IJ} &\rightarrow& m^{IJ} &+ \frac{1}{6} \Delta\C_{KLM}~ m^{KLMIJ}\\
m^{IJKLM} &\rightarrow& m^{IJKLM} &+ \frac{1}{2}
  \Delta\C_{NPQ}~m^{N;PQIJKLM}\\
m^{I;JKLMNPQ} &\rightarrow& m^{I;JKLMNPQ} &+ \frac{1}{2} \Delta\C_{RST}~
  m^{RSI;TJKLMNPQ}\\
m^{IJK;8} &\rightarrow& m^{IJK;8} &+ \frac{1}{6} \Delta\C_{LMN}~
  m^{LMNIJK;8}\\
m^{IJKLMN;8} &\rightarrow& m^{IJKLMN;8} &+ \frac{1}{2}
  \Delta\C_{PQR}~ m^{P;QRIJKLMN;8}\\
m^{1;8;8} &\rightarrow&m^{1;8;8}
\end{array}
\end{equation}
Here however, the $\C$ spectral flow turns out to be {\it non-integrable}.
Indeed, denoting by $\nabla^{IJK}$ the flow induced by the shift
$\C_{IJK} \rightarrow \C_{IJK} + \Delta\C_{IJK}$, we find
\begin{equation}
\label{comc}
\left[ \nabla^{IJK},\ \nabla^{LMN} \right] = 20 \nabla^{IJKLMN}
\end{equation}
where $\nabla^{IJKLMN}$ is the flow induced by the shift
$\E_{IJKLMN} \rightarrow \E_{IJKLMN} + \Delta\E_{IJKLMN}$:
\begin{equation}
\label{efl}
\begin{array}{llll}
m_I &\rightarrow& m_I &+ \frac{1}{5 !} \Delta\E_{JKLMNI}~ m^{JKLMN}\\
m^{IJ} &\rightarrow& m^{IJ} &+ \frac{1}{5!} \Delta\E_{KLMNPQ}~ m^{K;LMNPQIJ}\\
m^{IJKLM} &\rightarrow& m^{IJKLM} &+ \frac{1}{(3!)^2}
  \Delta\E_{NPQRST}~m^{NPQ;RSTIJKLM}\\
m^{I;JKLMNPQ} &\rightarrow& m^{I;JKLMNPQ} &+ \frac{1}{5!} \Delta\E_{RSTUVW}~
  m^{RSTUVI;WJKLMNPQ}\\
m^{IJK;8} &\rightarrow& m^{IJK;8} & +\frac{1}{5!} \Delta\E_{LMNPQR}
m^{L;MNPQRIJK;8} \\
m^{6;8} &\rightarrow& m^{6;8} & \\
m^{1;8;8} &\rightarrow& m^{1;8;8} &
\end{array}
\end{equation}
For $d \leq 7$, Eq.(\ref{comc}) is the only non-zero commutation relation,
while for $d=8$ the two flows  $\nabla^{IJK}$ and
$\nabla^{IJKLMN}$ close on a $\K_{1,8}$ flow. We shall, however, restrict
ourselves to the case $d\leq 7$ for simplicity. This non-commutativity
does not come as a surprise if one considers successive application of
the transformations that shift the values of the background fields $\C_{IJK}$
by integers. The reason is that a point in the homogeneous moduli space
$E_d(\Real)/K_d(\Real)$, where $K_d(\Real)$ is the maximal compact subgroup
of $E_d$, can be parametrized by a coset representative $g\in E_d(\Real)$~;
the latter can be represented according
to the Iwasawa decomposition
\begin{equation}
g\in E_d(\Real) = K_d(\Real) \cdot A_d(\Real) \cdot N_d(\Real)
\end{equation}
into compact $K_d$, abelian
$A_d$ and nilpotent $N_d$ factors; $N_d(\Real)$ can be thought of as
the group of upper triangular matrices with 1's on the diagonal,
whereas the compact factor is modded out in the quotient.
The gauge potentials (and the off-diagonal metric)
enter into the $N_d(\Real)$ factor, whereas
the (diagonal part of the) metric enters in the abelian factor $A_d(\Real)$.
Spectral flows act on $g$ from the right as elements of
$N_d(\Zint)$, and correspond to isometries of the scalar manifold.
They can be reabsorbed into a left action on the integer charge vector $m$,
so that the mass formula
\begin{equation}
\mathcal{M}^2= m^t\cdot g^t g \cdot m
\end{equation}
is invariant.
The nilpotent matrices $N_d(\Zint)$ exhibit commutation relations
graded by the distance away from
the diagonal, thus implying  non-commutativity for the spectral flows.
In the case at hand,  the
$\C_3,\E_6$ and $\K_{1;8}$ gauge potentials then
parametrize the first, second and third diagonal rows  respectively
above the main diagonal of $N_d(\Real)$.

The non-integrability can be evaded by combining the $\Delta \C_3$ shift
with a $\Delta \E_6$ shift,
\begin{equation}
\label{esh}
\frac{1}{5!} \Delta \E_{IJKLMN} =  \frac{1}{12} \C_{[IJK} \Delta \C_{LMN]}
\end{equation}
such that the resulting flow
\begin{equation}
\label{cfl}
\partial^{IJK} = \nabla^{IJK} - 10 C_{KLM} \nabla^{KLMIJK}
\end{equation}
becomes integrable.
The extra shift (\ref{esh}) is
invisible in the type IIA picture for
zero RR potentials since it does not contribute to the T-duality
spectral flow. We emphasize again that these terms are generated as a
consequence of integrability of the flow, which we take as a guide
for reconstructing the covariantized flow.
Note also that this flow does {\it not} preserve
the integer lattice of charges anymore, and consequently
does not deserve the name
of spectral flow; equivalently, it does not correspond to an isometry
of the scalar manifold $E_d/K_d$. Instead, the correct isometry is obtained
by accompanying the $\C_3$ shift (\ref{cfl}) by a compensating $\E_6$ shift
opposite to Eq.(\ref{esh}), and induces the true spectral flow (\ref{spcfl}).

The flow (\ref{cfl}) however allows us to integrate
the corresponding system of differential equations
\begin{equation}
\begin{array}{lllp{.5cm}lll}
\partial^{JKL} \tilde m_I  &  =  & \frac{1}{2}  \tilde m^{JK} \delta_I^L
& & \nabla^{JKLMNP} \tilde m_I  &  =  & \frac{1}{5!}  \tilde m^{JKLMN} \delta_I^P  \\
\partial^{KLM} \tilde m^{IJ} & = & \frac{1}{6} \tilde m^{KLMIJ}
&& \nabla^{KLMNPQ} \tilde m^{IJ} & = & \frac{1}{5!} \tilde m^{K;LMNPQIJ}  \\
\partial^{NPQ}  \tilde m^{IJKLM} & = &  \frac{1}{2} \tilde m^{N;PQIJKLM}
&& \nabla^{NPQRST}  \tilde m^{IJKLM} & = &  0 \\
\partial^{RST} \tilde m^{I;JKLMNPQ} & = & 0
& & \nabla^{RSTUVW} \tilde m^{I;JKLMNPQ} & = & 0
\end{array}
\end{equation}
to obtain the U-duality invariant mass formula
\enlargethispage{1cm}
for the flux multiplet in $d\leq 7$,
\begin{equation}
\label{mflux}
\M^2_F = \left( \tilde m_1 \right)^2 +
         \frac{1}{2!\ l_p^6} \left( \tilde m^{2} \right)^2 +
         \frac{1}{5!\ l_p^{12}} \left( \tilde m^{5} \right)^2 +
         \frac{1}{7!\ l_p^{18}}  \left( \tilde m^{1;7} \right)^2
\end{equation}
where the shifted charges read
\begin{equation}
\label{chshft}
\begin{array}{ll}
\tilde m_I   &= m_I  + \frac{1}{2} \C_{JKI} m^{JK}  +
\left( \frac{1}{4!} \C_{JKL} \C_{MNI} + \frac{1}{5!} \E_{JKLMNI}  \right) m^{JKLMN} \\
 & $\;\;\;\;\;$   + \left( \frac{1}{3! 4!}  \C_{JKL} \C_{MNP} \C_{QRI}
+ \frac{1}{2 \cdot 5!} \C_{JKL} \E_{MNPQRI} \right) m^{J;KLMNPQR} \\
\tilde m^{IJ}    &= m^{IJ} + \frac{1}{3!} \C_{KLM} m^{KLMIJ}
+ \left( \frac{1}{4!} \C_{KLM} \C_{NPQ} + \frac{1}{5!} \E_{KLMNPQ}  \right)
m^{K;LMNPQIJ} \\
\tilde m^{IJKLM}   &= m^{IJKLM}  + \frac{1}{2} \C_{NPQ}  m^{N;PQIJKLM}  \\
\tilde m^{I;JKLMNPQ}   &= m^{I;JKLMNPQ}
\end{array}
\end{equation}
This formula is written for the case $d=7$ and is invariant under $E_7(\Zint)$.
It reduces to the exact mass formulae in $d<7$ by simply
dropping the forms with more than $d$ antisymmetric indices.

As an illustration of the T-duality invariance, we display
the shift in the T-duality vector charge $m^{s1}$ implied by the
above equation:
\begin{equation}
\label{singshft}
\begin{split}
\tilde m^{s1} + \A_1 \tilde m^2 =& m^{s1}
  + \left[ \A_1 m^{2} +\left( \C_3 + \A_1 B_2 \right) m^{s4}
              +\left( \E_{s5} + \C_3 B_2 + \A_1 B_2 B_2 \right) m^{s;s6}
\right] \\
             &+ \left[ \A_1 \C_3 m^5 + \left( \E_6 + \C_3^2 + \A_1 \E_{s5}  +
\A_1 B_2 \C_3  \right) m^{1;s6} \right]\;.
\end{split}
\end{equation}
The first bracket in this expression precisely involves the tensor product
of  the {\it charge} spinor representation $S_B$ with the {\it RR moduli}
spinor representation. Indeed, the multiplet
$(\A, \C+\A B, \E + \C B + \A B^2 )$ 
transforms as a spinor multiplet, since it appears in the expansion
of the T-invariant D-brane coupling $e^{B+F} \R$ in powers of $F$.
The combination of moduli $( \A_1 \C_3 , \E_6 + \C_3^2 + \A_1 \E_{s5}  +
\A_1 B_2 \C_3   )$  on the other hand should transform as part of
a second order tensor under T-duality.

\subsection{The momentum multiplet}
Having obtained the full U-duality
invariant mass formula for the flux multiplet,
we now briefly discuss the case of the momentum multiplet.
As shown in Ref.\ \cite{egkr}, applying U-duality on
a state of mass $\M=R_l R_I / l_p^3$ generates masses
$$
\frac{R_l R_I R_J R_K R_L}{l_p^6},\
\frac{R_l R_I^2 R_J R_K R_L R_M R_N}{l_p^9},\
\frac{R_l R_I^2 R_J^2 R_K^2 R_L R_M R_N R_P}{l_p^{12}},\
\frac{R_l R_I^2 R_J^2 R_K^2 R_L^2 R_M^2 R_N^2 R_P}{l_p^{15}},\dots
$$
The dots stand for many extra contributions occurring when $d\ge 8$.
For simplicity, we shall restrict ourselves to $d\le 7$. The integer
charges corresponding to these states can be written as a set
of integer forms
$$n^1,\ n^4,\ n^{1;6},\ n^{3,7},\ n^{6,7}\ .$$
Decomposing these representations according to the
$g_s^2$ grading as in Eq.\ (\ref{gdec}), we find
that they combine under T-duality as
$$S=(n^s) \sp S_A=(n^1,n^{s3},n^{s;s5}) \sp
T=(n^4,n^{s;6},n^{1;s5},n^{s2;s6}),$$
$$S'_A=(n^{1;6},n^{3;s6},n^{s5;s6})\sp
S'=(m^{6;s6}).$$
The singlet $S$ exists in any dimension, while in $d-1=6$, $T$
contains an antisymmetric $SO(6,6)$ two-form and a singlet.
The spinor representation
$S'_A$ and the singlet $S'$ only exists in $d-1=6$.
Applying the same reasoning as for the flux multiplet,
we obtain the $E_6(\Zint)$-invariant mass formula for
the momentum multiplet in the case $d=6$:
\begin{equation}
\label{mmom}
\M^2_M = R_l^2 \left[ \frac{1}{l_p^6} \left(   \tilde n^1 \right)^2 +
     \frac{1}{l_p^{12}}     \left( \tilde n^{4} \right)^2 +
        \frac{1}{l_p^{18}} \left( \tilde n^{1;6} \right)^2 \right]
\end{equation}
where the shifted charges are given by
\begin{equation}
\label{sftm}
\begin{array}{ll}
\tilde m^{I}    &= m^{I} + \frac{1}{3!} \C_{JKL} m^{JKLI}
+ \left( \frac{1}{4!} \C_{JKL} \C_{MNP} + \frac{1}{5!} \E_{JKLMNP}  \right)
m^{J;KLMNPI} \\
\tilde m^{IJKL}   &= m^{IJKL}  + \frac{1}{2} \C_{MNP}  m^{M;NPIJKL}  \\
\tilde m^{I;JKLMNP}   &= m^{I;JKLMNP}
\end{array}
\end{equation}
As the overall factor $R_l^2$ in Eq. (\ref{mmom})
shows, the momentum multiplet describes
extended objects with one world-volume direction wrapped on the
longitudinal (light-like)
circle\footnote{Without mention of DLCQ, one could also understand
the momentum multiplet as the {\it multiplet of strings} of M-theory,
with tension $\mathcal{M}_M/R_l$.}.
States with $n^I$ charge correspond to membranes
wrapped on $R_l$ and a transverse radius $R_I$, and states with
$n^{IJKL}$ charge correspond to five-branes wrapped on four transverse
directions besides the longitudinal direction. The last charge
$n^{1;6}$ corresponds
to Taub-NUT gravitational monopoles. The mass formula (\ref{sftm})
can be extended to $d=7,8$ although the index structure soon becomes
intricate.

\subsection{Solitons and instantons}
As an illustration of the momentum multiplet
mass formula, we display the $d\le 5$ case,
where only $n^1$ and $n^4$ contribute:
\begin{equation}
\label{mmom5}
\begin{split}
\M^2_M =& \; \frac{R_l^2}{l_p^6}
 \left[ \left( n^I + \frac{1}{3!} n^{IJKL} C_{JKL} \right)
g_{IM} \left( n^M + \frac{1}{3!} n^{MNPQ} C_{NPQ} \right) \right. \\
&\left. \hspace{2cm}+ \frac{1}{4!\ l_p^6} n^{IJKL} g_{IM} g_{JN} g_{KP} g_{LQ} n^{MNPQ} \right] \;.
\end{split}
\end{equation}
This is precisely the U-duality invariant quantity obtained in the study
of instanton corrections to $R^4$ couplings in type II theories
\cite{inst,pk}, where it
was found that in order to obtain an $SO(5,5,\Zint)$-invariant result,
one should include, in addition to the D0-branes (described by $n^1$) and
the D2-branes (described by $n^{s3}$), extra states with a four-form charge
$n^4$ \cite{pk}\footnote{The charge
$n_s$ in Ref.\ \cite{pk} was associated to the
integer dual (in the sense of Poisson resummation) to the number of
D-branes bound together.}. It was further noticed that these states
would give $e^{-1/g_s^2}$ effects, which came as a surprise since
$T^4$ compactifications of type II string do not seem to allow
for NS five-brane instantons. In the present framework,
 $n^1$ and $n^4$ naturally appear
as membranes and five-branes wrapped on the longitudinal direction
in addition to one or four transverse directions, giving
solitons in the remaining six-dimensional theory. One should
therefore think of the non-perturbative threshold obtained in
Ref.\ \cite{pk} as a sum of {\it soliton loops} rather than of
instanton effects. This conclusion should however be taken with care,
since we have not been able to show that the
$SO(5,5,\Zint)$ Eisenstein series obtained from Eq.\ (\ref{mmom5})
contains the correct one-loop $R^4$ coupling.

\section{Gauge backgrounds in Matrix theory}

Gauge backgrounds of M-theory should have a counterpart as
couplings in the Matrix gauge theory.
In this Section, we will translate  the mass formulae of the M-theory
BPS states obtained in the previous Section
into the gauge theory language, and show that they arise
from topological couplings in the gauge theory. We will  determine
these couplings from D-brane analysis.

\subsection{BPS states of Matrix gauge theory}

As already emphasized in Ref.\ \cite{egkr}, the translation
from the M-theory mass to the light-cone energy, equated
to the energy in the Yang-Mills theory, differs
for the flux and momentum multiplets. The general formula
for bound states of flux and momenta states (\ie having
non zero values of both the $m$ and $n$ charges) reads
\begin{equation}
E_{YM} = \frac{\M_F^2}{P^+} + \sqrt{ \M_M^2 } \;
\end{equation}
where $\M_F$ and $\M_M$ are the masses of the flux and momentum multiplet
given in (\ref{mflux}) and (\ref{mmom})  respectively,
and $P^+=N/R_l$ is the quantized light-cone momentum.
Expressing $(l_p,R_s,g_{IJ})$ in terms of the Yang-Mills
parameters $(g^2, \tilde g_{IJ})$, and restricting
for simplicity to $d\le 7$, we obtain:
\begin{equation}
\label{eym}
\begin{split}
E_{YM}=&\frac{g^2}{N V_s}
\left[  \left( \tilde m^1 \right)^2 +
         \left(\frac{V_s}{g^2}\right)^2 \left( \tilde m_{2} \right)^2 +
         \left(\frac{V_s}{g^2}\right)^4 \left( \tilde m_{5} \right)^2 +
         \left(\frac{V_s}{g^2}\right)^6 \left( \tilde m_{1;7} \right)^2
\right]\\
&+\sqrt{  \left( \tilde n_1 \right)^2 +
         \left(\frac{V_s}{g^2}\right)^2 \left( \tilde n_{4} \right)^2 +
         \left(\frac{V_s}{g^2}\right)^4 \left( \tilde n_{1;6} \right)^2 +
         \left(\frac{V_s}{g^2}\right)^6 \left( \tilde n_{3;7} \right)^2 +
         \left(\frac{V_s}{g^2}\right)^8 \left( \tilde n_{6;7} \right)^2
} \; ,
\end{split}
\end{equation}
where the index contractions are now performed with the dual
metric $\tilde g_{IJ}= g^{IJ} l_p^6/R_s^2$. All upper indices
in the M-theory picture are turned into lower indices in
the Matrix gauge theory picture.
For $d\le 3$, Eq.\ (\ref{eym}) reduces to
\begin{equation}
\label{eym3}
\begin{split}
E_{YM}=&
\frac{g^2}{N V_s} \left( m^I + \frac{1}{2} \C^{IJK} m_{JK} \right)
\tg_{IL} \left( m^L + \frac{1}{2} \C^{LMN} m_{MN} \right) \\
&+\frac{V_s}{N g^2} \left( m_{IJ} \tg^{IK} \tg^{JL} m_{KL} \right)
+\sqrt{  n_I \tg^{IJ} n_J } \;.
\end{split}
\end{equation}
This includes the energy of the electric flux $m^I$
(\ie the momentum conjugate to $\int F_{0I}$) and the magnetic flux
$m_{IJ}=\int F_{IJ}$ in the diagonal Abelian subgroup of $U(N)$,
together with the energy of a massless excitation with quantized momentum
$n_I$. The shift of the electric flux $m^I$ in the presence of a
$\C_3$ gauge field background is the manifestation of the Witten
phenomenon \cite{wittenef} and indicates that the coupling of $\C_3$
to the gauge theory occurs through a topological term
$\int \C^{IJK} F_{0I} F_{JK}$.
Indeed, the only effect of such a coupling
is to shift the momentum conjugate to $\partial_0 A_I$ by a quantity
$\C^{IJK} \int F_{JK}$. In the next Subsection, we will derive the existence
of this coupling from the D-brane action.
 
When $d=4$, an extra charge $n_4$ appears in the momentum contribution,
which can be interpreted as the momentum along a (dynamically generated)
fifth dimension of radius $g^2$ \cite{rozali}.
We can indeed rewrite Eq.\ (\ref{eym})
in a U-duality ($Sl(5,\Zint)$)-invariant way as
\begin{equation}
E_{YM} = \frac{1}{N V_5} m^{AB} \tg_{AC} \tg_{BD} m^{CD} +
\sqrt{ n_A \tg^{AB} n_B } \;,
\end{equation}
where $A,B,\dots$, now run from 1 to 5, and $V_5=V_s g^2$ is the
volume of the five-dimensional torus. One may now interpret $m^{AB}$
as the quantized flux (in the diagonal Abelian group)
conjugate to a $U(N)$ two-form gauge field
$B_{AB}$ living on the 1+5 world volume. Note that the dependence of
$E_{YM}$ on the volume of the five-dimensional volume is through a
global factor $V_5^{-1/5}$. This agrees nicely with the scale invariance
of the conjectured 1+5-dimensional gauge theory \cite{rozali}.

\subsection{D-brane gauge couplings and Matrix theory}

The Matrix theory prescription
for M-theory compactifications may be recovered by viewing the DLC
light-like compactification as an infinitely boosted space-like
compactification described by weakly coupled type IIA string
theory\footnote{Subtleties may hide
behind this formal equivalence \cite{hp}.}.
The Matrix gauge theory  is then identified as the
gauge theory on the world-volume of $N$ D$d$ branes, obtained by
a maximal  T-duality from the $N$ D0-branes \cite{sen,seiberg}.
Whereas so far the prescription was only applied for compactifications with
vanishing gauge field expectation values, one may extend this argument
to find the couplings that these VEVs induce in the gauge theory, from
the well-known gauge couplings of D-branes.

The D$d$-brane T-dual to the $ N$ D0-brane interacts with the
RR fields through a topological Wess-Zumino term \cite{douglas}
\begin{equation}
\label{topo}
S_{\rm RR} =
  \int {\rm d} t \int_{\tilde{T}^{d}}
\STr e^{F+B} \wedge \R \;,
\end{equation}
where $\tilde{T}^d$ is the dual torus and the integral picks up
the contribution of $d+1$ forms in the integrand. $F$ is the $U(N)$ field
strength and
$\R =\sum_{p} \R^{(p)} $ is the total RR potential. The symmetrized  trace is
taken in the adjoint representation of $U(N)$ and will be omitted
henceforth.
The NS two-form $B$ couples to the Abelian diagonal part of the $U(N)$
field strength $F$; it would appear
after T-duality if we were considering DLCQ of
M-theory in the presence of a background value of $\C_{-IJ} $,
where the minus sign denotes the compact light-cone coordinate.
This case has been addressed recently in Ref.\cite{cds} and seems to require
drastic changes in the compactification prescription. We will therefore
restrict our attention to  $B=0$, in which case the metric on the dual torus
is the inverse of the M-theory metric.

For $d \leq 8$, the topological coupling truncates to
\begin{equation}
S_{\rm RR} =
\int {\rm d} t  \int_{\tilde{T}^{d}} \left[
 \R^{(d+1)}
 +  F  \R^{(d-1)}
 + \frac{1}{2} F^2  \R^{(d-3)}
 + \frac{1}{3!}F^3   \R^{(d-5)}
 + \frac{1}{4!}F ^4   \R^{(d-7)}  \right] \;,
\label{dba}
\end{equation}
where we have omitted the wedge products for notational simplicity.
The fields $\R^{(p)}$ are pulled back from the target space onto the
D$d$-brane world volume, with the embedding coordinates $X^{\m} (\sigma)$.
We will work in the static gauge in which the target space
coordinates of the torus coincide with the world-volume coordinates of
the D-brane.

The fields $\R^{(p)}$ are related by
a T-duality, on the  $d$ spatial  directions of the D$d$-brane world volume,
to the RR fields in the original D$0$-brane picture.
The action of this maximal T-duality on a $q \leq d+1$ RR form is
\begin{equation}
\R^{(q)}_{0 i_1 \ldots i_{q-1} }  \rightarrow \R^{(d+2-q)}_{0 i_q \ldots i_d } \sp
\R^{(q)}_{i_1 \ldots i_q }   \rightarrow \R^{(d-q)}_{i_{q+1} \ldots i_d } \;,
\end{equation}
where we have distinguished two cases, depending on whether or not
the RR field has a component in the time direction or not.
Ignoring for a moment the transverse fluctuations around the
D$d$-brane background, the Matrix model action in the D0-brane picture
is:
\begin{equation}
S_{\rm RR}  = \int {\rm d} t \int_{\tilde{T}^{d}}
\left[
 \R^{(1)}_0
 +  F_{0i}  \R^{(1)}_{i}
 +  F_{ij}  \R^{(3)}_{0 i j }
 + 
F_{0 i} F_{jk}  \R^{(3)}_{ijk}
 + 
F_{i j } F_{kl }  \R^{(5)}_{0 ijkl}
\right.  \end{equation}
$$
 + 
F_{0i } F_{jk} F_{lm}   \R^{(5)}_{ijklm}
 + 
F_{ij } F_{kl} F_{mn}   \R^{(7)}_{0ijklmn}
$$
$$
\left. + 
F_{0i } F_{jk} F_{lm} F_{nr}    \R^{(7)}_{ijklmnr}
 + 
F_{ij } F_{kl} F_{mn} F_{rs}   \R^{(9)}_{0ijklmnrs}
\right] \;,
$$
where  $\R^{(1)}= {\cal{A}}$ is the type IIA RR one-form,
$\R^{(3)}= {\cal{C}}$ is the RR three-form, etc. The
time component $\A_0$ can
be gauge transformed to zero since the time coordinate
is non-compact.
The type IIA gauge fields
arise under reduction of the M-theory metric and
(dual) gauge fields on the
space-like radius of radius $R_s$, as discussed below
Eq.\ (\ref{msa}). After a large boost of rapidity $\beta=1-(R_s/R_l)^2$,
this circle becomes quasi-lightlike and the metric takes
the form\footnote{The coordinates $x^-$ and $x^{i=1\dots d}$ are compact
variables with radius $R_l$ and $R_i$ respectively,
while $x^+$ and the spacetime coordinates $x^\mu$ are
non compact.}
\begin{equation}
ds^2 = dx^{+} \left(dx^{-} + \A_i dx^{i}\right) + dx^i g_{ij} dx^j\ .
\end{equation}
We can therefore identify $\A_i$ with $g_{+I}/R_l^2$, where
we promoted the string theory spatial index $i$ to the
M-theory transverse direction $I$. At the same time,
$\C_{ijk}$ is identified with the M-theory transverse three-form
$\C_{IJK}$, whereas the NS two-form $B_{ij}$ would turn into
$C_{-IJ}$, as already anticipated at the beginning of this section.
$\R^{(5)}_{0ijkl}$  turns into $\E_{-+IJKL}$,
while $\R^{(7)}_{ijklmnp}$
and
$\R^{(7)}_{0ijklmn}$ become the
components $\K_{-;-IJKLMNP}$ and $\K_{-;-+IJKLMN}$
of the M-theory $\K_{1;8}$ form in the DLC.
The nine-form
$\R^{(9)}$ is associated to a type IIA cosmological constant term and will
be discarded below.

Using these identifications we may then immediately
read off the coupling of the supersymmetric gauge theory to the
M-theory backgrounds:\footnote{As we noted below
Eq. (\ref{eym}), the lower indices of the
M-theory fields turn into upper indices in the Matrix gauge theory.}
\begin{equation}
\label{sm}
 S_{\rm Matrix} = \int {\rm d} t  \int_{\tilde{T}^{d}}
\left[
 F_{0I} g^{+I}
 +  F_{IJ}  \C^{+IJ}
 +  F_{0I} F_{JK}  \C^{IJK}
 +  F_{IJ} F_{KL }  \E^{-+IJKL}
\right.
\end{equation}
$$
\left.
\;\;\;\;\;\;\;
 +  F_{0I } F_{JK} F_{LM}   \E^{-IJKLM}
 +  F_{IJ } F_{KL} F_{MN}   \K^{-;-+IJKLMN}
+  F_{0I } F_{JK} F_{LM} F_{NP} \K^{-;-IJKLMNP}
\right] \;.
$$
The only term involving an eleven-dimensional Lorentz scalar, and therefore
a genuine modulus of M-theory compactification,  is the third term.
As a consequence we find that the expectation value of the three-form induces
the following topological coupling in the Matrix gauge theory :
\begin{equation}
S_{\C}  = \C^{IJK}~\int {\rm d } t \int_{\tilde{T}^d}
F_{0I} F_{JK}  \; ,
\end{equation}
as inferred from the gauge theory energy (\ref{eym3}).
This also agrees with the conjecture in  Ref.\ \cite{bcd}.
Some of the remaining terms in Eq. (\ref{sm}) were observed in
Ref.\ \cite{cds} and in the supermembrane context \cite{dewit}.

We next turn to the possible effects of terms containing transverse
fluctuations on the D$d$-brane. These will arise through the
expansion of the $q$-forms,
\begin{equation}
\R^{(q)} = \sum_{p=0}^q \R_{i_1 \ldots i_{q-p} \m_1 \ldots \m_{p} }
D_{i_{q-p+1}}  X^{\m_1} \wedge
\cdots \wedge D_{i_{q}} X^{\m_{p} } \;\;.
\end{equation}
For any non-zero value of $p$, these couplings will always involve
forms with at least one component in the spacetime directions,
transverse to the brane.
Since the maximal T-duality does not affect the transverse space,
the dual RR forms still involve transverse indices. Consequently, such terms
cannot generate couplings of the M-theory moduli to the gauge theory.

This reasoning does not touch  upon the couplings of the
$\E_6, \K_{1; 8}$ moduli
(related to $\C_3$ under U-dualities when $d\ge 6$)
to the Matrix gauge theory.
Such fields correspond to NS
gauge potentials in the D-brane picture, and are obviously not
incorporated in the topological coupling Eq.\ (\ref{topo}).
The partial picture obtained for $d\ge 6$ may well be related
to the difficulties in defining the Matrix gauge theory in
these dimensions.

\section{Nahm-type duality and eleven-dimensional Lorentz invariance}

In the last two Sections, we discussed the occurrence of the
$E_d (\Zint)$ U-duality symmetry both from the point of view of
M-theory and its Matrix gauge theory DLCQ description.
However, eleven-dimensional Lorentz invariance implies that this symmetry
should extend to an $E_{d+1}(\Zint)$ action on the M-theory BPS spectrum,
to which  we now turn.
 
As already noticed in Ref.\ \cite{egkr}, many of the states of
the flux multiplet, describing various branes wrapped on $k$ transverse
directions, have a counterpart in the momentum multiplet as the
same brane wrapping $k-1$ transverse directions and the IMF (or
light-cone) compact direction. Indeed, comparing the two mass
formulae (\ref{mflux}) and (\ref{mmom}), we see that we can
interpret the $E_d(\Zint)$ flux $m$ and momentum $n$ charges
as charges of a {\it flux} multiplet $M$ of $E_{d+1}(\Zint)$:
\begin{equation}
\begin{array}{lp{2cm}l}
m_1 = M_1  & &
m^{1;7} = M^{1;7} \sp n^{1;6}=M^{1;l6} \\
m^2 = M^2\sp n^1 = M^{l1} & &
m^{3,8} = M^{3;8} \sp n^{3,7}=M^{3;l7} \\
m^5 = M^5\sp n^4 = M^{l4} & &
m^{6,8} = M^{6,8} \sp n^{6,7}=M^{6;l7} \\
\end{array}
\end{equation}
where we now denote the light-cone direction by an index $l$ to avoid
confusion.
A notable exception is the transverse
KK state $m_1$ of the flux multiplet, with mass $1/R_I$,
which does not correspond to any
state with mass $1/R_{l}$ in the momentum multiplet. The reason
is clear: the longitudinal momentum is fixed in a given Matrix gauge
theory to equal the rank of the $U(N)$ gauge group. Following
the suggestion
in Ref.\ \cite{dlcq}, we regard
the tensor product of all gauge theories for all
values of $N$ as defining a M(eta) theory on which the
eleven-dimensional Lorentz symmetry is represented. $N$ would then
appear as an additional charge $M_{l}$ necessary to label the vacuum
of M(eta) theory. When $d\ge 6$, the singlet
\begin{equation}
N = M_l
\end{equation}
should be accompanied by
\begin{equation}
\label{rank}
N^{2;7}\equiv M^{l2;l7}\sp N^{6}\equiv M^{l;l6}\sp N^{5;7}\equiv M^{l5;l7}\sp
N^{1;7;7}\equiv M^{1;l7;l7}
\end{equation}
and, when $d=7$, by two extra singlets
\begin{equation}
N^{7}=M^{l;7}\sp N^{7;7}=M^{l;l7;l7}\ .
\end{equation}
The charges in Eq.\ (\ref{rank})
label a new U-duality multiplet that transforms as a {\bf 56} of $E_7(\Zint)$
(as is easily seen by dualizing the {\bf 6} and {\bf 7}
indices to $N^2,N_1,N^5,N^{1;7}$).
For $d=6$, it simply reduces to a singlet of $E_6(\Zint)$. We shall hereafter
refer to these new charges and $N$ as forming the (reducible)
{\it rank multiplet}. The dimension of the three U-duality multiplets
for $1  \leq d \leq 8$, as well as the U-duality group and the
dimension of the corresponding scalar manifold, are listed in the table below.
\vskip .2cm
\begin{equation*}
\hspace*{-.7cm}
\begin{array}{|c|c||c|c|c|c|c|c|c|c|}
\hline
\multicolumn{2}{|c||}{d} &  1  & 2 & 3      & 4 & 5 & 6   & 7 & 8 \\ \hline
\hline
\multicolumn{2}{|c||}{E_d(\Zint) } &
 1  & Sl(2)  & Sl(3) \times Sl(2)   &
Sl(5) & SO(5,5) &   E_6    & E_7  & E_8   \\
\hline
\multicolumn{2}{|c||}{\rm scalars} &
       1  & 3 &  7   & 14  & 25 & 42   & 70 & 128 \\
\hline \hline
{\rm Flux}   & \{m\}   & 1 & 3 & (3,2)  &10 &16 &27   &56 &248\\
{\rm Momentum} & \{n\} & 1  & 2 & (3,1)  &5  &10 &27   &133&3875\\
{\rm Rank}  &  \{ N \}  & 1 &  1 & 1      & 1 & 1 & 1+1 &56+1+1+1&\infty \\
\hline
{\rm Total}  & \{ M\}   & 3 & 6 & 10     &16 &27 &56   &248&\infty\\
\hline
\end{array}
\end{equation*}
\vskip .5cm
The extra generators of $E_{d+1}(\Zint)$ correspond to an extra
{\it Weyl} generator
exchanging the
light-cone direction with a chosen direction $I$ on $T^d$
($R_l \leftrightarrow R_I$ for the case of a rectangular torus), and
a {\it Borel} generator, corresponding to the spectral flow
$\C_{lJK} \rightarrow \C_{lJK} +1$ for some directions $J,K$
on the torus.

(i) As is obvious from the
derivation in Sec. 2.2 of Ref.\ \cite{egkr},
the addition of the Weyl transformation
$R_l \leftrightarrow R_{I}$ to the Weyl group of $E_d$
enhances the latter to ${\cal W}(E_{d+1})$.
Note in particular that for $d=8$, this is the (infinite) Weyl group
of the affine Lie algebra $E_9$, which implies the appearance
of an infinite set of multiplets in addition to the flux,
momentum and rank multiplet of $E_8(\Zint)$; for
$d=9$, this is the Weyl group of the hyperbolic algebra $E_{10}$.
We shall refrain from diving in these waters and restrict to $d\le 7$.
The action of this Weyl transformation on the parameters
(for rectangular tori) is by definition $R_l \leftrightarrow R_I$,
while leaving the other $R_J$'s and $l_p$ invariant.
In particular, the Newton constant in $11-(d+1)$ dimensions
\begin{equation}
\frac{1}{\kappa^2} = \frac{ V_R R_{l} }{l_p^9}
= R_{l}^{(d-7)/2} \frac{ V_s^{(d-5)/2}}{ g^{d-3} }
\end{equation}
is invariant under U-duality\footnote{This is
the same $E_d(\Zint)$-invariant combination as
appeared in Ref.\ \cite{egkr}, up to powers of $R_l$, which
anyway does not transform under $E_d(\Zint)$.}.
In terms of the M(atrix) theory, this means
\begin{equation}
g^2 \rightarrow \left(\frac{R_l}{R_I}\right)^{d-4}g^2 \sp
   s_I \rightarrow s_I \sp s_{J\ne I} \rightarrow
              \left(\frac{R_l}{R_I}\right) s_J \ .
\end{equation}
Note that the transformed parameters depend on the original ones
{\it and on} $R_l$. On the other hand, the only dependence
of the gauge theory on $R_l$
should be through a multiplicative factor in the Hamiltonian,
since $R_l$
can be rescaled by a Lorentz boost.  This leaves open
the question of how the M(eta) theory itself depends on $R_l$.
The action on the charges $M$ simply follows from the exchange of
the $I$ and $l$ indices. In terms of flux, momentum and rank charges,
this means
\begin{equation}
\label{z2nahm}
\begin{array}{lp{1cm}r}
\multicolumn{3}{c}{N \leftrightarrow m_I}\\
   n^{1} \leftrightarrow m^{I1}  &&
   N^{2;7} \leftrightarrow m^{I2,I7} \\
   n^{4} \leftrightarrow m^{I4}  &&
   N^{6} \leftrightarrow m^{I,I6} \\
   n^{1;6} \leftrightarrow m^{1;I6}  &&
   N^{5;7} \leftrightarrow m^{I5,I7}  \\
   n^{3;7} \leftrightarrow m^{3;I7}  &&
   N^{1;7;7} \leftrightarrow m^{1;I7;I7} \\
   n^{6;7} \leftrightarrow m^{6;I7}  &&
   N^{7;7} \leftrightarrow m^{I;I7,I7}\\
\multicolumn{3}{c}{N^7 \leftrightarrow n^{I;6}}\\
\end{array}
\end{equation}
In particular, the rank $N$ of the gauge group is exchanged with the
electric flux $m_I$, whereas the momenta are exchanged with
magnetic fluxes. This is reminiscent of Nahm duality,
relating (at the classical level) a $U(N)$ gauge theory
on $T^2$ with background flux $m$ to a $U(m)$ gauge theory on
the dual torus with background flux $N$ \cite{nahm}. There is, however,
no proof at this stage that this duality survives quantum
corrections and dimensional oxydation. This may eventually
be proved by a stringy argumentation.

(ii) The Borel generator $\C_{lJK}\rightarrow \C_{lJK}+
\Delta\C_{lJK}$ is
obtained from the usual $E_{d}(\Zint)$ shifts by conjugation
under Nahm-type duality. It is therefore not an independent generator,
but still gives a spectral flow on the BPS spectrum:
\begin{equation}
\label{shiftnahm}
\begin{array}{lcl}
N       &\rightarrow& N   + \Delta \C_{l2}~ m^2\\
m_1     &\rightarrow& m_1 + \Delta \C_{l2}~ n^{1}\\
m^2     &\rightarrow& m^2 + \Delta \C_{l2}~ n^4\\
m^5     &\rightarrow& m^5 + \Delta \C_{l2}~ n^{1;6}\\
m^{1;7} &\rightarrow& m^{1;7}+ \Delta \C_{l2}~ n^{3;7}\ ,\\
\end{array}
\end{equation}
the other charges being non-affected.
In particular, this implies that states with negative $N$
need to be incorporated in the M(eta) theory if it is
to be $E_{d+1}(\Zint)$-invariant. This is somewhat
surprising since the DLC quantization selects $N>0$,
and seems to require a revision both of the interpretation
of $N$ as the rank of a gauge theory and of the relation
between $N$ and the light-cone momentum $P^+$\footnote{See
Ref.\ \cite{rey} for a discussion of DLCQ with negative
light-cone momentum.}.
The resolution of the first point
may come from the conjecture, made in Ref.\ \cite{cds},
that the Yang-Mills
theory should be replaced,
in the presence of a background value for $\C_{lIJ}$,
by a gauge theory on
a non-commutative torus. Rather than being a liability, this may
actually turn into an asset by carrying the non-commutative geometry
constructions into the quantum  realm.

{\bfseries Acknowledgements : }
This work was supported in part by the EEC
under the TMR contract ERBFMRX-CT96-0090,
the Israel Academy of Sciences and Humanities -
Centers of Excellence Programme, and the American-Israel Bi-National
Science Foundation.

\end{document}